\documentclass[12pt]{article}
\usepackage{amsthm,amsmath,amsfonts,amssymb}
\usepackage{amsmath,amsfonts,amsthm}
\newcommand{\R}{{\mathord{\mathbb R}}}

\newcommand{\C}{{\mathord{\mathbb C}}}

\newcommand{\KK}{\mathcal{K}}
\newcommand{\HH}{\mathcal{H}}

\newcommand{\FF}{\mathcal{F}}
\newcommand{\CC}{\mathcal{C}}

\newcommand{\hh}{\mathfrak{h}}

\newcommand{\ben}{\begin{displaymath}}
\newcommand{\een}{\end{displaymath}}
\newcommand{\beqn}{\begin{equation}}
\newcommand{\eeqn}{\end{equation}}
\newcommand{\beqna}{\begin{eqnarray*}}
\newcommand{\eeqna}{\end{eqnarray*}}

\newtheorem{lemma}{Lemma}
\newtheorem{theorem}[lemma]{Theorem}

\newtheorem{corollary}[lemma]{Corollary}
\newtheorem{definition}[lemma]{Definition}

\title{On the self-adjointness  and domain of  Pauli-Fierz type Hamiltonians}

\author{\vspace{5pt}
D. Hasler\footnote{E-mail: dh8ud@virginia.edu.}
\
and I. Herbst\footnote{E-mail: iwh@virginia.edu.}
 \\
\vspace{-2pt}
\small{Department of Mathematics, University of Virginia,} \\
\small{Charlottesville, VA 22904-4137, USA}}

\date{ }

\begin{document}

\maketitle

\begin{abstract} We prove a general theorem about the self-adjointness and
domain of Pauli-Fierz type Hamiltonians. Our proof is based on
commutator arguments which allow us to treat fields with non-commuting
components. As a corollary it follows that the
domain of the Hamiltonian of non-relativistic QED with Coulomb
interactions is independent of the coupling constant.
% {\tt selfalt12.tex}
\end{abstract}

\section{Introduction}

Pauli-Fierz Hamiltonians are at the foundation of a  mathematically consistent description
of non-relativistic quantum mechanical matter
interacting with the quantized electromagnetic field.  For a Hamilton operator to
describe a unitary dynamics it
must be self-adjoint. Thus the question of self-adjointness is
intimately related to physics. Knowing the domain of
self-adjointness turns out to be of technical relevance for proving
various properties about the Hamiltonian.
%, such as existence and absence of ground states \cite{losmiyspo:low,hasher:abs}.

In this paper we prove a general theorem stating that the
domains of Pauli-Fierz type Hamiltonians are independent of  the coupling strength.
%We also give estimates wich may be useful.
Our proof is based on
elementary commutator arguments, which allow us to treat fields of
general form. Thus our theorem does  not require the components
of the fields
to commute (see Theorem \ref{thm:mainnoext}). In such a case
functional integral methods are typically
not applicable. As a corollary, we show that the domain of the
Hamiltonian of non-relativistic QED with Coulomb interactions is
independent of the coupling constant. Such  a result has been
obtained previously using functional integral methods, see
\cite{sim:the,glijaf:qua,feffrogra:sta,hir:ess,hir:sel}.
However, an operator theoretic proof has sofar been lacking in the
literature. % (see \cite{spo:dyn}, p. 164).

The paper is organized as follows. First, we introduce definitions
and collect some elementary properties in lemmas. Although these
properties are well known, a proof is given in the Appendix for the
convenience of
the reader. The Hamiltonian of the interacting system is realized as
the self-adjoint operator associated to a semi-bounded quadratic
form. In a first step we show using a commutator argument that the
domain of the free Hamiltonian is an operator core for the
interacting Hamiltonian (see Lemma \ref{lem:core}).
In a second step we show using operator
inequalities
that the free
Hamiltonian is operator bounded by the interacting Hamiltonian on a
suitable core for the free Hamiltonian (see Lemma \ref{eq:last2}).
Our result then follows as an application of the closed graph theorem.

\section{Model and Statement of Result}

Consider the Hilbert space $L^2(\R^n)$. For a measurable
function $f: \R^n \to \C$, we define the  multiplication operator
$ M_f \varphi := f \varphi$ for all $\varphi$ in the domain
$D(M_f) = \{ \varphi \in L^2(\R^n) | f\varphi \in L^2(\R^n) \}$. If
$f$ is real valued, then $M_f$ is self-adjoint. Let $p_j$ be the
operator defined  by, $p_j \varphi := -i\partial_j \varphi := -i (
\partial_j \varphi)_{\rm dist}$,  for  $\varphi$ in the domain
$$D(p_j) := \{ \psi \in L^2(\R^n) | (\partial_j \psi )_{\rm
dist} \in L^2(\R^n)\} ,$$ where $(\cdot )_{\rm dist}$ stands for the
distributional derivative and $\partial_j$ stands for the partial
derivative with respect to the  $j$-th coordinate in  $\R^n$. The
Laplacian is defined by $- \Delta := p^2  :=  \sum_{j=1}^n p_j^2$  with domain
$ D(p^2) :=  H^2(\R^n)$.
The operators $p_j$ and
$p^2$ are self-adjoint on their domains.

In this paragraph, we review some standard conventions about
tensor products, which can be found for example in
\cite{reesim:fun}.  The algebraic tensor product $V \otimes W$ of
the vector
spaces $V$ and $W$ consists of all finite linear combinations of
vectors of the form $\varphi \otimes \eta$ with $\varphi \in V$ and
$\eta \in W$. For $\HH$ and $\KK$ two Hilbert spaces the tensor
product of Hilbert spaces is the closure of the algebraic tensor
product of $\HH$ and $\KK$ in the topology induced by the inner
product.
We   adopt the standard convention that $V \otimes W$ denotes the
tensor product of Hilbert spaces if  $V$ and $W$ are Hilbert spaces;
if $V$ or $W$ is  a non complete inner product space
then $V \otimes W$ denotes the algebraic tensor product. For $A$ and
$B$   closed operators in the Hilbert spaces $\HH$ and $\KK$,
respectively, we denote by $A \otimes 1$  the closure of
$A \otimes 1 \upharpoonright D(A)\otimes \KK$  and by $ 1 \otimes B $
the closure of  $1 \otimes  B
 \upharpoonright \HH \otimes D(B)$.
%associate the operators in $\HH \otimes \KK$,
%\begin{align}  \label{eq:tensordef}  A \otimes 1  :=
%\overline{ A \otimes 1 \upharpoonright D(A)
%\otimes \KK}  \  , \quad    1 \otimes B :=  \overline{1 \otimes  B
% \upharpoonright \HH \otimes D(B)} \; .
%\end{align}
If $A$ is essentially self-adjoint on $D_a$,
then $A \otimes 1$ is essentially self-adjoint on $D_a \otimes \KK$. An analogous statement holds for
$1 \otimes B$.
%If $A$ and $B$ are essentially self-adjoint on $D_a$ and $D_b$,
%then $A \otimes 1$ and $1 \otimes B$
%are essentially self-adjoint on $D_a \otimes \KK$ and $\HH \otimes D_b$, respectively.
For notational
convenience, the operators $A \otimes 1$ and $B \otimes 1$
are written as  $A$ and  $B$, respectively.
 No confusion should arise, since it should be
clear from the context in which space the operator acts. By associativity and
bilinearity of the tensor product, the above definitions, conventions,
and properties generalize in a straight forward way to multiple
tensor products, \cite{reesim:fun}.

Let $\hh$ be a separable complex Hilbert space and let $\otimes^n
\hh = \hh \otimes \hh \otimes \cdots \otimes \hh$ denote the
$n$-fold tensor product of $\hh$ with itself. We define the Hilbert
spaces
$$
\mathcal{F} :=  \bigoplus_{n=0}^\infty  \FF_n \, , \quad  \FF_0 :=
\C \, , \quad  \FF_n :=  S_n (\otimes^n \hh) \;  , \  \ n \geq 1 \; ,
$$
where $S_n$ denotes the orthogonal projection onto totally symmetric
tensors, i.e., the projection satisfying $ S_n(\varphi_1 \otimes
\varphi_2 \otimes \cdots \otimes \varphi_n ) = \frac{1}{n !}
\sum_{\sigma \in \mathcal{S}_n} \varphi_{\sigma(1)} \otimes
\varphi_{\sigma(2)} \otimes \cdots \otimes \varphi_{\sigma(n)}$,
with $\mathcal{S}_n$ being the set of permutations of the numbers
$1$ through $n$. By definition, a vector $\psi \in \FF$ is a
sequence $(\psi_{(n)})_{n\geq 0}$ of vectors $\psi_{(n)} \in \FF_n$
such that its norm $(\sum_{n=0}^\infty \| \psi_{(n)} \|^2)^{1/2}$ is
finite.  Let $\Omega = ( 1 , 0, 0 , ... )$, and let
$$
\FF_{\rm fin} = \{ \psi \in \FF | \ \psi_{(n)} = 0 \ {\rm except \
for \ finitely \ many} \ n \}
$$
denote the subspace  consisting of states containing only finitely
many ``particles''.  Let $A$ be a self-adjoint operator on $\hh$ with domain $D(A)$.
The
second quantization $d\Gamma(A)$ is an operator in $\FF$ defined
as
follows. Let $A_{(n)}$ be the closure of
\begin{eqnarray*}
\left( A \otimes 1 \otimes ... \otimes 1 + 1 \otimes A \otimes 1
\otimes ... \otimes 1   +   ... + 1 \otimes ... \otimes 1 \otimes A \right)
 \upharpoonright S_n ( \otimes^n D(A)) \; .
\end{eqnarray*}
Define
$
(d \Gamma(A) \psi)_{(n)} = A_{(n)} \psi_{(n)}
$
for all $\psi$ in the domain
 $ D(d \Gamma(A)) := \{ \psi \in \FF| \psi_{(n)} \in D(A_{(n)}) \ , \
\sum_{n=0}^\infty \| A_{(n)} \psi_{(n)} \|^2 < \infty \}$. It
follows from the definition that  $d\Gamma(A)$ is self-adjoint. The number operator is defined
by $N = d\Gamma(1)$.
For each $h \in \hh$ we define the creation operator $a^*(h)$ by
$$
a^*(h) \varphi  =  (n+1)^{1/2} S_{n+1} h \otimes \varphi \quad ,
\quad \forall \varphi \in \FF_n \; ,
$$
and extend $a^*(h)$  to be an operator in $\FF$ by taking the
closure. Let $a(h)$ be the adjoint of $a^*(h)$. The annihilation
operator $a(h)$ acts on $\FF_0$ as the  zero operator and on vectors $S_n( \varphi_1
\otimes \cdots  \otimes  \varphi_n )\in \FF_n$, with $n \geq 1$,  as
$$
a(h) S_n (\varphi_1 \otimes \varphi_2 \otimes \cdots \otimes \varphi_n )
=  n^{-1/2} S_{n-1}  \sum_{i=1}^n (h, \varphi_i)  \varphi_1 \otimes \cdots
\varphi_{i-1} \otimes \varphi_{i+1} \otimes \cdots  \otimes
\varphi_n \; .
$$
For $h \in \hh$, we introduce the field operator on $\FF_{\rm fin}$
$$
\widehat{\phi}(h) =
2^{-1/2} ( a(h) + a^*(h) )  \; \; .
$$
This operator is symmetric, and hence closable. Let $\phi(h)$ denote the
closure of $\widehat{\phi}(h)$.

We shall henceforth assume
that $\hh = L^2(\R^d; \C^p)$ and that  $\omega :\R^n \to [0,\infty)$ is a measurable function which is
a.e. nonzero.
%\vspace{0.5cm} \noindent {\bf Hypothesis ($\boldsymbol{\omega}$).}
%Let $\omega :\R^n \to [0,\infty)$ be  a measurable function which is
%a.e. nonzero.
%\vspace{0.5cm}
The field energy, defined by,
$$
H_f = d \Gamma(M_\omega) \;
$$
is self-adjoint in $\FF$. It is notationally convenient to
define  the Hilbert space
$\hh_\omega :=  \{ h \in \hh | \| h \|_\omega < \infty \}$   with norm
$\| h \|_\omega := (\| h \|^2 + \| h /\sqrt{\omega} \|^2 )^{1/2}$.
In the next lemma we collect some basic and well known properties. A
proof of the lemma can be found in the Appendix.

\begin{lemma} \label{lem:fockrel} The following statements hold.
%For $h \in \hh$, $\FF_{\rm fin}$ is an operator core for $\phi(h)$.
\begin{itemize}
\item[(a)]  For $ g,h \in
\hh$, % $[\phi(g), \phi(h) ] = i {\rm Im}(g,h)$ on $\FF_{\rm fin}$.
\begin{eqnarray} \label{eq:comm1}
 [\phi(g), \phi(h) ] = i {\rm Im}(g,h) \quad {\it on} \ \
\FF_{\rm fin} \; .
\end{eqnarray}
\item[(b)]  If
$h \in \hh_\omega$, then $D(H_f^{1/2}) \subset D(\phi(h))$ and
\begin{eqnarray} \label{eq:basicineq1}
\| \phi(h) (H_f + 1 )^{-1/2}  \| &\leq& 2^{1/2} \| h  \|_\omega  \; .
\label{eq:basicineq33}
\end{eqnarray}
If  $g, h \in \hh_\omega$, then $D(H_f) \subset D(\phi(g)\phi(h))$
and
\begin{eqnarray} \label{eq:basicineq2}
\| \phi(g) \phi(h)  (H_f + 1 )^{-1} \| &\leq& 4 \| g \|_\omega \| h
\|_\omega   \; .
\end{eqnarray}
\item[(c)] If $h, \omega h \in \hh$, then
\begin{eqnarray} \label{eq:comm4}
[H_f , \phi(h) ] = - i \phi (i \omega h )  \quad {\it on} \ \ \FF_{\rm fin} \cap D(H_f) \; .
\end{eqnarray}
\end{itemize}
\end{lemma}

Now we will
extend the above definition to the tensor product $ \HH = L^2(\R^n)
\otimes \FF$ of Hilbert spaces. We will use the natural isomorphism of
Hilbert spaces,
\begin{eqnarray*} %\label{eq:isohilbert}
\HH := L^2(\R^n) \otimes \FF \cong L^2(\R^n ; \FF) \; ,
\end{eqnarray*}
and we introduce the space
$$
\HH_{\rm fin} = \{ \psi \in \HH | \ \psi_{(n)} = 0 \ , \  {\rm except \
for \ finitely \ many} \ n \} \; .
$$
Let $L^\infty(\R^d ; \hh)$ and $L^\infty(\R^d ; \hh_\omega)$ denote
the Banach spaces of measurable functions from $\R^d$ to  $\hh$ and $\hh_\omega$
with norms $\| G \|_\infty := {\rm ess \, sup}_{x \in \R^d}
\| G(x)\|$ and $\| G \|_{\omega,\infty} := {\rm ess \, sup}_{x \in \R^d} \|
G(x)\|_{\omega}$, respectively.
For $G \in L^\infty(\R^n ; \hh)$ define
 $\widehat{\Phi}(G)$  for $\psi \in \HH_{\rm fin}$ by
$$
(\widehat{\Phi}(G) \psi)(x) = {\phi}(G(x)) \psi(x) \; .
$$
Note that $\widehat{\Phi}(G)$ is a symmetric operator and hence closable.
Let $\Phi(G)$ denote the closure of $\widehat{\Phi}(G)$.

\vspace{0.5cm} \noindent {\bf Remark.} Although not needed for the
proof of the theorem, we note that  $\phi(f)$ and $\Phi(G)$ are
essentially self-adjoint on $\FF_{\rm fin}$ and $\HH_{\rm fin}$, respectively.  This can be shown
using, for example, Nelson's analytic vector theorem, see \cite{brarob:ope}.

\begin{lemma} \label{lem:basicineq0}
Let  $G \in L^\infty(\R^d ; \hh_\omega)$. Then   $D(H_f^{1/2})
\subset D(\Phi(G))$ and
\begin{eqnarray} \label{eq:basicineq3}
\| \Phi (G) (H_f + 1 )^{-1/2}  \| &\leq&  2^{1/2} \| G  \|_{\omega,
\infty}
\end{eqnarray}
\end{lemma}
\begin{proof} Follows from inequality \eqref{eq:basicineq1}.
\end{proof}

\begin{lemma}  \label{lem:closed} Let $(G_i)_{i=1}^n \subset L^\infty(\R^n;
\hh_\omega)$, and $A_j :=  \Phi(G_j)$. Then the quadratic form
\begin{eqnarray} \label{eq:formq}
q(\varphi,\psi) := \sum_{j=1}^n ( (p_j + A_j )\varphi, (p_j + A_j)
\psi ) + ( H_f^{1/2} \varphi, H_f^{1/2} \psi ) \; ,
\end{eqnarray}
defined on the form domain $Q(q) := \bigcap_{i} D(p_i)\cap
D(H_f^{1/2})$ is nonnegative and closed.
\end{lemma}

The proof of this Lemma is given in the Appendix.

\begin{definition} For $(G_j)_{j=1}^n \subset L^2(\R^n ;
\hh_\omega)$,  let $T_A$  be  the unique self-adjoint
operator associated to the quadratic form \eqref{eq:formq}.
For $G_j \in L^2(\R^n ; \hh)$, we will set $A_j :=  \Phi(G_j)$.
% and $p_{A,j} := p_j + A_j$.

\end{definition}

 \vspace{0.5cm} \noindent {\bf Remark.}
By the first representation theorem for quadratic forms, $T_A$ is characterized as follows:
\begin{eqnarray} ( T_A \varphi, \psi ) = q(\varphi,\psi) \quad , \quad \forall
\psi \in Q(q) \; ,
\end{eqnarray}
for all $\varphi$ in the domain $ D(T_A) = \{ \varphi
\in Q(q) | \exists \eta \in \HH , \, \forall \psi \in C , \
q(\varphi,\psi ) = (\eta , \psi ) \} $, where $C$ is any form core
for $q$.

\vspace{0.5cm}

\begin{definition} $G \in L^\infty(\R^n ;  \hh)$ is said to be weakly
$\partial_j$-differentiable
if there
is a $K \in L^\infty(\R^n; \hh)$ such that for all $v \in \hh$ and
all $f \in C_0^\infty(\R^n)$
$$
\int \partial_j f(x) ( v , G(x) ) dx = - \int f(x) ( v , K(x) ) dx
\; ;
$$
in that case we write $\partial_j G = K$.
\end{definition}

\vspace{0.4cm}

\noindent {\bf Hypothesis ($\boldsymbol{G}$).}  $(G_j)_{j=1}^n
\subset
 L^\infty(\R^n;\hh_\omega)$ is a collection of functions such that $G_j$ is weakly
 $\partial_j$-differentiable and   $   \omega G_j  , \sum_{l=1}^n \partial_l G_l \in L^\infty(\R^n;
\hh_\omega)$.

\vspace{0.4cm}

 We will adopt  standard
conventions for the sum and the composition of two operators: $D(S+R) = D(S) \cap D(R)$ and $D(SR)
=\{ \psi \in D(R) | R \psi \in D(S) \}$.  Since $p^2$ and $H_f$ are commuting  positive operators,
$p^2 +H_f$ is self-adjoint  on the domain $D(p^2 + H_f) = D(p^2) \cap D(H_f)$.

\begin{theorem} \label{thm:mainnoext} Let Hypothesis %($\omega$) and
($G$) hold. Then
$T_A$ is essentially self-adjoint on
any operator core for $p^2 + H_f$ and
$
D(T_A) = D(p^2 + H_f)  .
$
\end{theorem}

The next theorem relates $T_A$ with
 a natural definition.  By $(p+A)^2$ we denote the operator sum $\sum_{j} (p_j + A_j)^2$.
Thus by definition
\begin{eqnarray*}
\lefteqn{ \varphi \in D((p+A)^2) } \\
& \Longleftrightarrow&  \varphi \in D(p_j) \cap D(A_j) \ \ {\rm and} \ \ (p_j \varphi  + A_j\varphi) \in D(p_j) \cap D(A_j)
\ , \ \ \forall j = 1, .. , n \; ;
\end{eqnarray*}
and $D( (p + A)^2 + H_f) = D((p+A)^2) \cap D(H_f)$.

\begin{theorem} \label{thm:main22} Suppose %Hypothesis ($\omega$)  holds and
 $G_j \in L^\infty(\R^n;\hh_\omega)$ is weakly
 $\partial_j$-differentiable and
$\partial_j G_j \in L^\infty(\R^n ; \hh_\omega)$ for $j=1,...,n$.
Then  $ D(p^2 + H_f) \subset D(T_A) \cap  D((p+A)^2 + H_f)$.
Furthermore,
$$(p+A)^2 \varphi + H_f \varphi = T_A \varphi  \; , \quad {\it if}  \quad   \varphi \in D(p^2 + H_f) \; .
$$
\end{theorem}

\vspace{0.5cm}

\section{Applications}

Let  $\HH = \otimes^n ( L^2( \R^3) \otimes \C^2 )) \cong
L^2(\R^{3n} ; \otimes^n \C^{2}) $ be the Hilbert space, describing $n$
spin-$\frac{1}{2}$ particles. Let $x_j$ denote the coordinate of the $j$-th particle
having mass $m_j>0$, and let $x = (x_1, ... , x_n) \in \R^{3n}$. Let
\begin{eqnarray*}
\boldsymbol{\sigma}_{j,a} = 1 \otimes \cdots 1 \otimes \sigma_{a} \otimes 1
\cdots \otimes 1
%{j-{\rm th} \ {\rm position}}
\; ,
\end{eqnarray*}
where  $\sigma_a$, the $a$-th Pauli matrix, acts on the $j$-th
factor of $\otimes^n \C^2$. Let $\hh = L^2(\R^3 ; \C^2 )$, and
let $\varepsilon(1,k)$ and $\varepsilon(2,k)$ be normalized vectors
in $\C^3$ depending measurably on $k/|k|$ such that $(\varepsilon(i,k) , k )
=0$, for $i=1,2$ and $( \varepsilon(1,k) ,  \varepsilon(2,k)) = 0$.
Let $\omega(k) = \sqrt{ m^2_{\rm ph} + k^2 }$ for some $m_{\rm ph}
\geq 0$. Let $\rho(k)$ be a function such that $\rho/\omega ,
\sqrt{\omega} \rho \in L^2(\R^3)$.
For $a=1,2,3$ and $j=1, ..., n$,
let
\begin{align}
& [G_{j,a}(x)](k,\lambda) = \frac{\rho(k)}{\sqrt{ \omega(k)}} e^{-i k \cdot x_j}
\varepsilon_a(\lambda,k)
 , \ \   [E_{j,a}(x)](k,\lambda) =
\frac{-i \rho(k)}{\sqrt{ \omega(k)}} e^{ -i k  \cdot x_j} ( k \wedge
\varepsilon(\lambda,k))_a \; , \label{eq:coupling}
\end{align}
and
$
A_{j,a} = \Phi(G_{j,a})$ and $B_{j,a} =  \Phi(E_{j,a})$.
Let $V_c : \R^{3n} \to \R$ be a function which is infinitesimally
bounded with respect to $-\Delta := p^2 $. For example this is the
case, if for  $c_{j,l}, z_{j,J} \in \R$ and $(R_J)_{J=1}^M  \subset \R^3$,
$$
V_c = \sum_{j \neq l} \frac{c_{j,l}}{|x_j - x_l|} + \sum_{j=1}^n
\sum_{J=1}^M  \frac{z_{j,J}}{| x_j - R_J |} \; .
$$
We want to point out that one usually imposes the constraint that
$\rho(k) = \overline{\rho(-k)}$, which is not needed for the corollary below to hold.
Moreover, note that $[A_{a,l}, A_{b,j}] = 0$ is satisfied
only if  $|\rho(k) | = |\rho(-k)|$ (see Lemma  \ref{lem:extrel}).

\begin{corollary} \label{thm:main2} The operator
\begin{eqnarray} \label{eq:paulifierz}
\sum_{j}   \frac{1}{2 m_j}( - i \nabla_j - e_j A_{j} )^2  + H_f +
\sum_j \frac{e_j}{2 m_j} \boldsymbol{\sigma}_j \cdot B_j  + V_c \; ,
\end{eqnarray}
with $e_j \in \R$, is well defined on $D(\sum_j
\frac{- \Delta_j }{2m_j}  + H_f)$. It is  self-adjoint with this domain,
essentially self-adjoint on any core for
$\sum_j \frac{- \Delta_j }{2m_j}  + H_f$, and bounded from below.
\end{corollary}

Clearly the same result holds if we restrict the operators to  subspaces taking into account certain
particle statistics. The statement of this corollary has been
previously obtained using functional integral methods, \cite{hir:sel}.

\begin{proof} After rescaling the particle coordinates and the  functions \eqref{eq:coupling},
we can assume that
$m_j =1$ and $e_j = -1$. The
$G_{j,a}$ (possibly a rescaled version thereof) satisfy the assumptions of
Theorems~\ref{thm:mainnoext} and \ref{thm:main22}. Thus
by Theorem~\ref{thm:main22}, $(p+A)^2 + H_f$ is well defined on $D(p^2 + H_f)$. Moreover,
for $\varphi \in D(p^2 + H_f)$, we have
$(p+A)^2\varphi + H_f\varphi = T_A\varphi$.
By Theorem~\ref{thm:mainnoext}, $D(p^2 + H_f) = D(T_A)$ and therefore
$T_A$ is  $ p^2 +
H_f$ bounded. Since $ \boldsymbol{\sigma}_{j} \cdot B_{j}  $ and $V_c$ are
infinitesimally small with respect to $p^2 + H_f$, the claim
follows now from Kato's Theorem.
\end{proof}

\section{Proofs}

%Throughout this section we assume that Hypothesis ($\omega$) holds.
% We use the standard
%conventions  $D(S+R) = D(S) \cap D(R)$ and $D(SR)
%=\{ \psi \in D(R) | R \psi \in D(S) \}$.
We use the convention that $[S,R]$ stands for the operator
$SR-RS$ defined on the  domain $D([S,R]) = \{ \psi \in D(R) \cap
D(S) | S \psi \in D(R) , \ R \psi \in D(S) \}$.

\begin{lemma} \label{lem:extrel} The following statements hold.
\begin{itemize}
\item[(a)]
 For $F,G \in L^\infty(\R^n ; \hh)$,
$$
 [ \Phi(F) , \Phi(G) ] = i {\rm Im} ( F , G)_\hh    \quad  {\it on } \ \HH_{\rm fin} ,
$$
where the right hand side is a multiplication operator and the inner product is  taken in
$\hh$.
\item[(b)] If $F,G \in L^\infty(\R^n ; \hh_\omega)$, then $D(H_f^{1/2}) \subset D(\Phi(G))$ and
$$
\| \Phi (G) (H_f + 1 )^{-1/2}  \| \leq  2^{1/2} \| G  \|_{\omega, \infty}
$$
moreover,  $ D(H_f) \subset D(\Phi(F)\Phi(G)) $ and
$$
\| \Phi (F) \Phi(G) ( H_f + 1 )^{-1}  \| \leq  4 \| F  \|_{\omega,
\infty} \| G \|_{\omega, \infty} \; .
$$
\item[(c)]
For $G, \omega G  \in L^\infty(\R^n ; \hh)$,
$$
[ H_f , \Phi(G) ] =
 - i \Phi(i \omega G) \ {\it on } \ \HH_{\rm
fin} \cap D(H_f) \; .
$$
\item[(d)]  Let  $G \in L^\infty(\R^n ; \hh)$ be weakly $\partial_j$-differentiable.
Then $\Phi(G)$ leaves $
\HH_{\rm fin} \cap D(p_j)$  invariant and
$$
[p_j , \Phi(G) ] =  - i \Phi( \partial_j G)  , \quad {\it on} \
 \HH_{\rm fin} \cap D(p_j)  \; .
$$
\end{itemize}
\end{lemma}
\begin{proof} All statements up to and including (c) follow
 directly from the definition and
corresponding statements  in Lemma \ref{lem:fockrel}.
(d) Follows from Lemma \ref{lem:leibniz} in the Appendix.
\end{proof}

In the proof, we will use certain   commutator identities which can
be easily verified on a suitable core, which we shall now introduce.
Let
 $$
\mathcal{C} := C_0^\infty(\R^n) \otimes  \left(
 \coprod_{n=0}^\infty S_n (\otimes^n C^\omega )
 \right) \; ,
$$
where  $C^\omega$ denotes the set of functions $f$ in $L^2(\R^d;
\C^p)$ with ${\rm supp} f \subset  \bigcup_{m=0}^\infty \{ k | \omega(k) \leq m \}$,
and $\amalg_{n=0}^\infty S_n (\otimes^n C^\omega)$ denotes
the set of all sequences $(\psi_{(n)})_{n =0}^\infty$ such that
$\psi_{(n)} \in S_n (\otimes^n C^\omega)$ and $\psi_{(n)} =0$ for
all but finitely many $n$. Note that $\CC \subset \bigcap_{m=1}^\infty D(H_f^m)$.
If $G_j \in L^\infty(\R^n; \hh)$ is weakly $\partial_j$-differentiable, we have
by Lemma \ref{lem:extrel} (d),
\begin{eqnarray} \label{eq:pAsquared}
\CC \subset    D(p_j^2) \cap D(A_j p_j) \cap D(p_j A_j ) \cap D( A_j^2)
%\subset D( (p + A)^2 )
\; .
\end{eqnarray}

\vspace{0.5cm}

\begin{lemma} Let $(G_j)_{j=1}^n \subset L^\infty(\R^n ; \hh_\omega)$.  The set
$\mathcal{C}$ is a form core for $q$.
\end{lemma}

\begin{proof}
 By definition we have to show that
$\mathcal{C}$ is dense in $(Q(q), \| \cdot
\|_+)$, where
\begin{eqnarray} \label{eq:+1norm2}
 \| \varphi \|_{+1}^2 :=
\| \varphi \|^2 + \sum_{j}\|(p_j + A_j )\varphi\|^2 + \| H_f^{1/2}
\varphi \|^2 \; .
\end{eqnarray}
For $\psi \in Q(q)$, there exists a sequence
$(\psi_n)_{n=0}^\infty
\subset \mathcal{C}$ such that $ \psi_n \to \psi$, $p_j \psi_n \to
p_j \psi$, and $H_f^{1/2} \psi_n \to H_f^{1/2} \psi$.
 This and Lemma  \ref{lem:extrel} (b) imply that $(p_j + A_j )\psi_n
\to (p_j + A_j )\psi$. Thus $\mathcal{C}$ is dense in $(Q(q), \|
\cdot \|_+)$.
\end{proof}

Part  (c) of the next lemma immediately implies Theorem \ref{thm:main22}.
Parts (a),(b) and (d) will be used to prove Theorem \ref{thm:mainnoext}.

\begin{lemma}  \label{lem:core} Suppose for $j=1,...,n$, $G_j \in L^\infty(\R^n ; \hh)$ is weakly
$\partial_j$-differentiable. Then following statements are true.
\begin{itemize}
\item[(a)]   For all $\varphi \in \CC$,
% \label{eq:realization}
$T_A \varphi = (p + A)^2 \varphi  + H_f \varphi$.
\item[(b)] Let   $G_j, \sum_l \partial_l G_l \in L^\infty(\R^n ; \hh_\omega)$  for all $j=1,...,n$.  Then $D(p^2 + H_f) \subset D(T_A)$.
\item[(c)] Let $G_j, \partial_j G_j \in L^\infty(\R^n ; \hh_\omega)$  for all $j=1,...,n$. Then
$D(p^2 + H_f) \subset D((p+A)^2) \cap D(T_A)$ and  for all $\varphi \in D(p^2 + H_f)$,  $T_A \varphi = (p + A)^2 \varphi  + H_f \varphi$
.
\item[(d)] If Hypothesis ($G$) holds, the set  $D(p^2 + H_f)$ is an operator core for $T_A$.

\end{itemize}
\end{lemma}

\begin{proof}
\noindent
(a). Using \eqref{eq:pAsquared}, we see that for $\varphi , \psi \in \CC$,
$q(\varphi , \psi ) = \sum_j ( (p + A )^2_j
\varphi , \psi ) + ( H_f \varphi , \psi)$. This shows (a).

\noindent
(b).
Let $\varphi \in D(p^2 + H_f)$. By definition $\varphi \in D(p^2) \cap D(H_f)$. By  Lemma
\ref{lem:extrel}, $\varphi \in D(A_l^2)$.
Since $\| H_f^{1/2} p_l \varphi \|^2 \leq \| H_f \varphi\|^2 + \| p_l^2
\varphi \|^2$, we have $p_l \varphi  \in D(A_l)$.
Thus it follows that for $\psi \in \CC$,
\begin{eqnarray} \label{eq:4july}
q(\varphi, \psi ) = \sum_j \left\{ ( p_j^2 \varphi , \psi ) + (A_j p_j \varphi , \psi ) + ( A_j^2 \varphi , \psi) + ( \varphi, A_j p_j \psi)
\right\}  + (H_f \varphi , \psi) \; .
\end{eqnarray}
Now using  Lemma \ref{lem:extrel}   (d), we see that the summation over the last term in the sum
yields,
\begin{eqnarray*}
\sum_j ( \varphi , A_j p_j \psi ) &=& ( \varphi ,  i \Phi(\Sigma_j \partial_j G_j )\psi )   +  \sum_j ( \varphi , p_j A_j  \psi ) \\
&=&  ( - i \Phi(\Sigma_j \partial_j G_j )\varphi ,   \psi )
+  \sum_j ( A_j p_j  \varphi ,   \psi ) \; .
\end{eqnarray*}
Thus, there exists an $\eta \in \HH$, such that for all $\psi \in \CC$,
$q(\varphi , \psi ) = (\eta , \psi)$. This shows (b).

\noindent(c). In view of (b) and \eqref{eq:4july}, we only need to show
$A_j \varphi \in D(p_j)$.
By   Lemma \ref{lem:extrel} (d), for
$\varphi_n = \chi_{[0,n]}(N) \varphi$ with
 $\chi_{[0,n]}$ denoting the characteristic function of the set $[0,n]$,
$$
p_j A_j \varphi_n = - i \phi(  \partial_j G_j) \varphi_n + A_j p_j
\varphi_n \; .
$$
Since the limit of the right hand side exists and $A_j \varphi_n$ converges,
it follows that $A_j \varphi \in D(p_j)$.

\noindent
(d).
 Let $\alpha
> 0$.  For notational compactness, we set
 $
R_\alpha := ( \alpha H_f + 1 )^{-1}$ and
$ \Pi_j := \Phi( i \omega G_j )$.
Moreover, observe that
$D(|p|) = \cap_j D(p_j)$.

\vspace{0.5cm}

 \noindent \underline{Step 1:} For $\varphi \in D(T_A)$, and for all $\psi \in \mathcal{C}$,
\begin{eqnarray} \label{eq:23rdjunesecond}
q( R_\alpha \varphi, \psi)  = q(\varphi, R_\alpha \psi) + (F_\alpha \varphi , \psi )  + 2 \sum_j (E_{\alpha,j}
(p_j + A_j) \varphi, \psi ) \; ,
\end{eqnarray}
where   $E_{\alpha,j}$ and $F_\alpha$  are  bounded operators
defined by
\begin{eqnarray*}
E_{\alpha,j} &=& - i R_\alpha \alpha \Pi_j R_\alpha \; , \\
 F_\alpha & =&  - \sum_j 2 \alpha^2 R_\alpha \left( \Pi_j R_\alpha \right)^2
 -   \alpha R_\alpha^2  \sum_j ( \omega G_j , G_j )_\hh
 + i [ R_\alpha , \Phi( \Sigma_j \partial_j G_j ) ] \; .
\end{eqnarray*}

\vspace{0.5cm} \noindent To show this,  let $\varphi \in D(T_A)$ and
$\psi \in \mathcal{C}$. By definition $\varphi \in D(|p|)$. Since
$R_\alpha$ is bounded and acts on a different factor
of the tensor product it leaves $D(|p|)$ invariant. It follows that
$R_\alpha \varphi \in Q(q)$. By the definition of the
quadratic form,
\begin{eqnarray}
 q(R_\alpha \varphi,\psi)  = \sum_j \left(  \varphi, R_\alpha  (p_j + A_j)^2
\psi \right) + ( H_f^{1/2}  \varphi, H_f^{1/2} R_\alpha \psi ) \; . \label{eq:23rdjune}
\end{eqnarray}
We write  the summand in the first expression on the right as
\begin{eqnarray*}
 \left(  \varphi, R_\alpha  (p_j + A_j)^2
\psi \right)
&=&  \left(  \varphi, [R_\alpha , (p_j + A_j)^2] \psi
\right) +  \left(  \varphi, (p_j + A_j)^2  R_\alpha    \psi \right) \\
&=&  \left(  \varphi, [R_\alpha , (p_j + A_j)^2] \psi
\right) +  \left( (p_j + A_j ) \varphi, (p_j + A_j) R_\alpha \psi \right)
\end{eqnarray*}
Inserting this into \eqref{eq:23rdjune}  we find
$$
q( R_\alpha \varphi, \psi)  = q(\varphi, R_\alpha \psi) +
  \sum_j \left(  \varphi, [ R_\alpha  , (p_j + A_j)^2
] \psi \right) \; .
$$
We  calculate the commutator
\begin{eqnarray*}
\lefteqn{  \sum_j \left(  \varphi, [ R_\alpha , (p_j + A_j)^2
] \psi \right) } \\
&=&   \sum_j \left\{ \left(  \varphi, \left[[R_\alpha ,  A_j], p_j
+A_j \right]  \psi \right) + \left(  \varphi, 2 ( p_j + A_j)
[R_\alpha ,  A_j] \psi \right) \right\}
\\
&=&  \sum_j \left\{  \left(  \varphi, \left[[R_\alpha ,  A_j], A_j  \right]  \psi \right)
+ (\left( \varphi , \left[R_\alpha, [ A_j , p_j ] \right] \psi \right)  -
\left(2 ( p_j + A_j)  \varphi, E_{\alpha,j}  \psi \right) \right\} \\
&=& ( F_\alpha \varphi , \psi ) +   2 \sum_j (E_{\alpha,j}
(p_j + A_j) \varphi, \psi ) \; ,
\end{eqnarray*}
where we used that
on $\mathcal{H}_{\rm fin}$,
$$
E_{\alpha,j} = [ A_j , R_\alpha ] \quad , \quad \left[[R_\alpha ,  A_j], A_j  \right]  =  -
 2 \alpha^2 R_\alpha \left( \Pi_j R_\alpha \right)^2
 -   \alpha R_\alpha^2  ( \omega G_j , G_j )_\hh \; .
$$

\vspace{0.5cm}

 \noindent \underline{Step 2:} For all $\varphi \in D(T_A)$, $R_\alpha \varphi  \in D(T_A)$
 and
 $
\lim_{\alpha \downarrow 0} T_A R_\alpha \varphi  = T_A
\varphi
 $.

\vspace{0.5cm} \noindent From Eq. \eqref{eq:23rdjunesecond}, it follows
that  $R_\alpha \varphi  \in D(T_A)$ and  that
\begin{equation} \label{eq:1}
T_A R_\alpha \varphi = R_\alpha  T_A \varphi +
F_\alpha  \varphi  + 2 \sum_j E_{\alpha,j} (p_j + A_j) \varphi  \; .
\end{equation}
By the spectral theorem $s - \lim_{\alpha \downarrow 0} R_\alpha = 1 $.
Using the estimate
\begin{eqnarray*}
 \| \alpha^{1/2} \Pi_j (\alpha H_f + 1 )^{-1/2}\|   \leq  {\rm max}(1 , \alpha^{1/2} ) \| \Pi_j (H_f + 1
)^{-1/2} \|  \; ,
\end{eqnarray*}
we see that $E_{\alpha,j}$ and the first term of $F_\alpha$ converge to 0 for $\alpha \downarrow 0$.
Moreover,
\begin{align*}
%&  R_\alpha \Phi(\Sigma_j \partial_j G_j ) \varphi  \rightarrow \Phi(\Sigma_j \partial_j G_j ) \varphi \\
&  \Phi(\Sigma_j \partial_j G_j )  R_\alpha \varphi  =
 \Phi(\Sigma_j \partial_j G_j ) (H_f + 1 )^{-1/2}  R_\alpha  ( H_f  + 1 )^{1/2} \varphi \rightarrow
\Phi(\Sigma_j \partial_j G_j ) \varphi \; ,
\end{align*}
as  $\alpha \downarrow0$, which implies  $\lim_{\alpha \downarrow 0} [ R_\alpha,  \Phi(\Sigma_j \partial_j G_j ) ] \varphi = 0$.
Thus the right hand side of
Eq. \eqref{eq:1} converges for $\alpha \downarrow 0$ to $T_A \varphi$.

\vspace{0.5cm}

 \noindent \underline{Step 3:} For $\varphi \in D(T_A)$, and $\alpha >
 0$,
 $ R_\alpha \varphi \in D(p^2) \cap D(H_f)$.

\vspace{0.5cm} \noindent It is clear that $R_\alpha
\varphi \in D(H_f)$. Let $\varphi \in D(T_A)$. Then by
\eqref{eq:23rdjunesecond}
 there exits an
$\eta \in \HH$, such that for all $\psi \in \mathcal{C}$,
\begin{eqnarray*}
( \eta , \psi  )  &=& \sum_j ( (p_j + A_j) R_\alpha  \varphi, (p_j + A_j ) \psi
)  \\
&=&  \sum_j \left\{ ( p_j R_\alpha \varphi , p_j \psi) + ( A_j^2 R_\alpha \varphi, \psi)
+ ( p_j R_\alpha \varphi , A_j \psi) + ( A_j R_\alpha \varphi , p_j \psi) \right\} \;
\end{eqnarray*}
Furthermore, using $\sum_j ( A_j R_\alpha \varphi , p_j \psi) = ( R_\alpha \varphi ,
 i \Phi(\Sigma_j \partial_j G_j )\psi)    +  \sum_j (R_\alpha \varphi , p_j A_j \psi)$,
$\varphi~\in~D(|p|)$, and   Lemma \ref{lem:extrel},
we see that  there exists an
$\eta_1 \in \HH$, such that
$$
\sum_j ( p_j R_\alpha \varphi, p_j \psi ) = (\eta_1
,\psi) \quad , \quad \forall \psi \in \mathcal{C} \; .
$$
This implies $R_\alpha \varphi \in D(p^2)$, since
$\mathcal{C}$
is a form core for $p^2$.
\end{proof}

\begin{lemma}  \label{eq:last2} Let Hypothesis ($G$) hold. Then
there exists  constants $C_1, C_2$ such that for all $\varphi  \in \CC$,
\begin{eqnarray} \label{eq:2nd}
 \| ( p^2 + H_f ) \varphi  \|^2 \leq  C_1
 \| ( ( p + A)^2 + H_f )\varphi \|^2  + C_2 \| \varphi \|^2 \; .
\end{eqnarray}
\end{lemma}

\begin{proof}
The proof will be based on the relations given in  Lemma
\ref{lem:extrel}.
First observe that
\begin{eqnarray} \label{eq:2nd2}
\| (p^2 + H_f ) \varphi \|^2   \leq  2 \| p^2 \varphi \|^2 +
2 \|  H_f \varphi \|^2    \ ,
\quad \forall \varphi \in \CC \; .
\end{eqnarray}
The lemma will  follow as a direct consequence of
Inequality  \eqref{eq:2nd2} and Steps 1 and 2, below.

\vspace{0.5cm}

 \noindent \underline{Step 1:} There exist constants $c_1, c_2, c_3 $ such
that
$$
\| p^2 \varphi \|^2 \leq c_1 \| (  (p + A)^2  \varphi \|^2   +
c_2 \| H_f  \varphi \|^2   + c_3 \| \varphi \|^2 \ ,
\quad \forall \varphi \in \CC \; .
$$

We have
\begin{eqnarray*}
\|  p^2 \varphi \|^2
%&&= \left( ( p + A - A ) \cdot ( p + A - A ) \right)^2 \\
&&=  \|  (  ( p + A )^2  - A  \cdot ( p + A ) -  ( p + A)  \cdot A +
A^2 ) \varphi \|^2 \\
&&\leq 3 \| ( p + A )^2 \varphi \|^2  + 3 \|(  A  \cdot ( p + A
) +  ( p + A) \cdot A  ) \varphi  \|^2 + 3 \| A^2 \varphi \|^2 \; ,
\end{eqnarray*}
writing $A^2 = A \cdot A$. We estimate the middle term using the notation
$[ p  , A ] := \sum_j [p_j  , A_j ]$,
\begin{eqnarray}
 \| ( A  \cdot ( p + A ) +  ( p + A) \cdot A ) \varphi \|^2
&=& \|  ( 2 A \cdot (p + A)
+ [ p  ,  A ] ) \varphi \|^2   \nonumber \\
 &\leq&  8   \| A  \cdot ( p + A) )\varphi \|^2 +
 2 \| [p , A ] \varphi \|^2 \; . \label{eq:today33}
\end{eqnarray}
The second term on the last line is estimated using
 $ \|[ p ,  A ]\varphi \|^2 \leq C \| ( H_f  + 1 )^{1/2} \varphi \|^2$. Here and below $C$ denotes a
 constant which
 may change from one inequality to the next.
The first term in \eqref{eq:today33} is estimated as follows:
\begin{eqnarray*}
  \| A \cdot  ( p + A ) \varphi \|^2   \leq C \sum_j \| A_j ( p_j + A_j) \varphi \|^2
\leq C \sum_j \| ( H_f + 1 )^{1/2} ( p_j + A_j ) \varphi \|^2 \; .
  \end{eqnarray*}
Further, using a commutator
\begin{eqnarray*}
 \sum_j \| ( H_f + 1 )^{1/2} ( p_j + A_j ) \varphi \|^2
&=&
\sum_j ( (p_j + A_j) \varphi , ( H_f + 1 ) (p_j + A_j) \varphi ) \\
&=&  \sum_j  \left( ( ( p_j + A_j)^2 \varphi , ( H_f + 1 ) \varphi ) +
( ( p + A)_j \varphi , [ H_f , A_j ] \varphi ) \right)  \\
& \leq & C (  \| (p + A)^2 \varphi \|^2 +    \sum_j \| (p_j + A_j) \varphi \|^2     +
  \| ( H_f + 1 ) \varphi \|^2 ) \\
& \leq &  C (  \| (p + A)^2 \varphi \|^2 +      \| ( H_f + 1 ) \varphi \|^2 ) \; .
  \end{eqnarray*}
Collecting the above estimates yields Step 1.

\vspace{0.5cm}

 \noindent \underline{Step 2:} There exists two constants $C_1$ and $C_2$ such
that
\begin{eqnarray*}
\|(p + A)^2 \varphi \|^2  + \| H_f \varphi \|^2  \leq
C_1 \| ( ( p + A)^2 + H_f )\varphi \|^2 + C_2 \| \varphi \|^2 \ ,
\quad \forall \varphi \in \CC \; .
\end{eqnarray*}

\vspace{0.5cm}

%Let $a>0$, using \eqref{eq:basicineq2}  and a
Calculating a double commutator,
we see that
\begin{eqnarray}
 \lefteqn{  \frac{1}{2} \| H_f   \varphi \|^2    +  ( H_f \varphi , ( p + A)^2 \varphi ) +
(( p + A)^2 \varphi , H_f \varphi)   }  \nonumber \\
&=&   \frac{1}{2}  \| H_f   \varphi \|^2  +
\sum_j  2  ( (p_j +A_j) \varphi ,  H_f ( p_j +A_j) \varphi)
 \nonumber \\
 &&  +  \sum_j \left( (\varphi, [ A_j, [
A_j , H_f ]]  \varphi) +   ([A_j , p_j ]\varphi,   H_f    \varphi) +
( H_f \varphi , [ A_j , p_j ] \varphi )   \right) \nonumber  \\
&\geq& \frac{1}{4} \| H_f \varphi \|^2 - C   \| (H_f + 1 )^{1/2} \varphi \|^2     \nonumber \\
&\geq&  - b \| \varphi \|^2 \; ,  \nonumber % \label{eq:last1}
\end{eqnarray}
 for some $b$. Step 2 follows from this.

\end{proof}
\vspace{0.5cm} \noindent {\it Proof of Theorem \ref{thm:mainnoext}.}
By Lemma  \ref{lem:core} (b), we know the inclusion
$ D(p^2 + H_f) \subset D(T_A)$. From the closed graph
theorem it follows that $T_A$ is $p^2 + H_f$ bounded. This,
Lemma  \ref{lem:core} (d), and the fact that $\CC$ is an operator core for
$p^2  + H_f$, imply that $\CC$ is an operator core for $T_A$.
From this, Lemma \ref{lem:core} (a), and Inequality  \eqref{eq:2nd} we conclude that
$D(T_A) \subset D(p^2 + H_f)$.
The statement about the core holds for any closed operators having equal domain.
 \qed

\vspace{0.5cm}

\section*{Appendix}

 \noindent {\it Proof of Lemma
\ref{lem:fockrel}.}
\noindent
(a). Relation \eqref{eq:comm1}
follows from the following relations on $\FF_{\rm fin}$,
$[a(f),a(g)] = 0$, $[a^*(f), a^*(g)]=0$,  and $[a(f), a^*(g)]
=(f,g)$, for all $f,g \in \hh$.

\noindent
(b). We will use the natural isomorphism $\otimes^n \hh \cong L^2((\R^d \times \C^p )^n)$.
% and let $L^2_s((\R^d \times \C^p)^n) := U \FF_n$.
We set $K_n = (k_1, \lambda_1 ,  \ldots , k_n , \lambda_n ) \in
(\R^d \times \C^p )^n$ and write $\int d K_n$ for $\sum_{\lambda_1,
... \lambda_n =1}^p \int dk_1 ... dk_n$. For $\psi \in \FF_{\rm fin}$ and $f \in \hh_\omega$,
%$D(H_f^{1/2})$.
%Then
\begin{align}
\| a(f) \psi \|^2 &= \sum_{n=0}^\infty \int  \left| (n+1)^{1/2} \sum_{\lambda_1} \int
\overline{f}(k_1) \omega(k_1)^{-1/2} \omega(k_1)^{1/2}
\psi_{n+1}(k_1,\lambda_1,K_n ) dk_1 \right|^2  dK_n  \nonumber \\
&\leq   \| f /\sqrt{\omega} \|^2 \sum_{n=0}^\infty  \int \sum_{\lambda_1} \int (n+1)
\omega(k_1) |  \psi_{n+1}(k_1,\lambda_1,K_n )|^2  dk_1
dK_n  \nonumber \\
& = \| f /\sqrt{\omega} \|^2 ( \psi , H_f \psi ) \; .
\label{eq:comm1proof}
\end{align}
Thus $D(H_f^{1/2}) \subset D(a(f))$. For $\varphi \in \FF_{\rm
fin}$,
\begin{equation*} \label{eq:comm1proof2}
\| a^*(f) \varphi \|^2 = ( a^*(f)\varphi , a^*(f) \varphi) = \| f
\|^2 \| \varphi \|^2 + \| a(f) \varphi \|^2 \; .
\end{equation*}
By this and \eqref{eq:comm1proof} %, and since $a^*(f)$ is closed,
we find
$D(H_f^{1/2}) \subset D(a^*(f))$ and
$$
\| a^*(f) \psi \|^2 \leq \| f\|^2 \| \psi \|^2 + \| f/\sqrt{\omega}
\|^2 \| H_f^{1/2} \psi \|^2 \; .
$$
If $\psi \in \FF_{\rm fin}$
%$D(H_f)$
and $f,g \in \hh_\omega$, then with $c_n := (n+1)(n+2)$,
\begin{eqnarray}
\lefteqn{ \| a(f) a(g) \psi \|^2 }  \nonumber \\
&=& \sum_{n=0}^\infty \int \left|c_n^{1/2}  \sum_{\lambda_1,\lambda_2} \int
\overline{f}(k_1,\lambda_1) \overline{g}(k_2,\lambda_2)
\psi_{n+2}(k_1,\lambda_1,k_2,\lambda_2, K_n ) dk_1 dk_2   \right|^2  dK_n  \nonumber  \nonumber \\
&\leq&   \left\|\frac{ f}{ \sqrt{\omega}} \right\|^2 \left\| \frac{g }{\sqrt{\omega}} \right\|^2
\sum_{n=0}^\infty \int \sum_{\lambda_1,\lambda_2} \int c_n \omega(k_1) \omega(k_2) |
\psi_{n+2}(k_1,\lambda_1,k_2, \lambda_2,  K_n )|^2 dk_1 dk_2
dK_n  \nonumber \\
& \leq& \| f /\sqrt{\omega} \|^2  \| g /\sqrt{\omega} \|^2  \| H_f
\psi \|^2  \; . \label{eq:22}
\end{eqnarray}
Now using the commutation relations, linearity, and  the triangle inequality
 we can reduce Inequalities  \eqref{eq:basicineq1} and \eqref{eq:basicineq2}
to the estimates \eqref{eq:comm1proof} and \eqref{eq:22}.

\noindent
(c). This follows from the identities $[H_f, a(f) ] = - a(\omega f)$ and
$[H_f, a^*(f) ] = a^*(\omega f)$ on $\FF_{\rm fin} \cap D(H_f) $, which in turn
follow from the definition.
 \qed

\vspace{0.5cm}

 \noindent {\it Proof of Lemma  \ref{lem:closed}.} By
Lemma~\ref{lem:basicineq0}, the right hand side of \eqref{eq:formq}
is well defined. By
definition $q$ is closed if and only if $Q(q)$ is complete
under the norm  \eqref{eq:+1norm2}.
We
write $(a_n)$ as a shorthand notation for the sequence
$(a_n)_{n=0}^\infty$. Let $(\varphi_n) \subset Q(q)$ be a Cauchy
sequence with respect to the norm $\| \cdot \|_{+1}$, see
\eqref{eq:+1norm2}. We
 see that the sequences
\begin{eqnarray}  \label{eq:cauchyseq}
(\varphi_n), \  (H_f^{1/2} \varphi_n), \ ((p_j + A_j) \varphi_n), \
j=1,2,3,
\end{eqnarray}
are  Cauchy sequences in $\HH$. Since $ (H_f^{1/2} \varphi_n)$
is Cauchy in $\HH$, it follows from Lemma~\ref{lem:basicineq0} that
$(A_j \varphi_n)$ is also Cauchy in $\HH$. Hence also
%By this and the Cauchy
%property of $((p_j + A_j) \varphi_n)$ it follows  that
$(p_j
\varphi_n)$ is Cauchy in $\HH$. Since $p_j$ and $H_f$ are closed, it
follows that the limit $\varphi = \lim_{n \to \infty} \varphi_n$
 is in the domain of $Q(q)$.
We
conclude $\| \varphi - \varphi_n \|_{+1} \to 0$ as $n$ tends to
infinity. \qed

\vspace{0.5cm}

\begin{lemma}  \label{lem:leibniz}
Assume $G \in L^\infty(\R^n ; \hh)$ is weakly $\partial_j$-differentiable.
Then  for $\psi \in  D(p_j) \cap \HH_{\rm fin}$,  $\Phi(
G) \psi $ is in the domain of $p_j$ and
$$
p_j \Phi(G) \psi = - i \Phi( \partial_j G) \psi + \Phi(G) p_j \psi
\; .
$$

\end{lemma}
\begin{proof}
Suppose
\begin{eqnarray*}
\psi_1 = f_1 \otimes \xi_1 \ \ , & & \quad \xi_1 =  a^*(h_1) \cdots
a^*(h_M)\Omega  \; , \\
\psi_2 = f_2 \otimes \xi_2 \ \ , & & \quad \xi_2 = a^*(g_1) \cdots
a^*(g_N)  \Omega \; ,
\end{eqnarray*}
with $f_j \in C_0^\infty(\R^n)$ and $g_i,h_i \in \hh$. Then
\begin{eqnarray*}
 (i p_j \psi_1 , \Phi(G) \psi_2 ) &&=
( i p_j f_1 \otimes \xi_1 , \Phi(G) f_2 \otimes \xi_2 ) \\
&& = \int (\partial_ j f_1(x)) \left[ \left( \xi_1 , 2^{-1/2} (
a^*(G(x)) + a(G(x)) \xi_2 \right) f_2 (x) \right] dx  \\
&& = 2^{-1/2} \int \Big[ \partial_j f_1(x) \sum_l (\xi_1 , ( G(x) ,
g_l )
a^*(g_1) \cdots a^*(\widehat{g}_{l}) \cdots \Omega ) f_2(x)  \\
&& +   \sum_l   \partial_j f_1(x) ( ( G(x) , h_l) a^*(h_1) \cdots
a^*(\widehat{h}_{l})  \cdots \Omega , \xi_2 ) f_2(x)  \Big] dx \; ,
\end{eqnarray*}
where $a^{*}(\widehat{h}_l)$ stands for the omission of the term
$a^{*}({h}_l)$. We find,
\begin{eqnarray*}
 (i p_j \psi_1 ,  \Phi(G) \psi_2 )
&& = - 2^{-1/2} \int f_1(x) \Big[ \sum_{l} (\xi_1 , ( g_l ,
\partial_j G(x) ) a^*(g_1) \cdots a^*(\widehat{g}_{l})  \cdots
\Omega ) f_2(x)  \\
&& + \sum_{l} (\xi_1, (g_l, G(x)) a^*(g_1) \cdots
a^*(\widehat{g}_{l})  \cdots
\Omega ) \partial_j f_2 (x)  \\
&& + \sum_{l} ( ( h_l, \partial_j G(x) ) a^*(h_1) \cdots
a^*(\widehat{h}_{l})  \cdots \Omega , \xi_2 ) f_2(x) \\
&& +  \sum_{l}  ( ( h_l,  G(x) ) a^*(h_1) \cdots a^*(\widehat{h}_{l})
 \cdots \Omega , \xi_2 ) \partial_j f_2(x) \Big] dx \\
&& = (\psi_1 , - \Phi(\partial_j G ) \psi_2 ) + (\psi_1 , \Phi(G) (
- i p_j ) \psi_2) \; .
\end{eqnarray*}
 Since linear combinations of vectors of the form
$\psi_1$ constitute a core for $p_j$ and $p_j$ is self-adjoint we
find $\Phi(G) \psi_2 \in D(p_j)$ and
$$
p_j ( \Phi(G) \psi_2 ) = - i \Phi( \partial_j G) \psi_2 + \Phi(G)
p_j \psi_2 \; .
$$
This equation now follows for any $\psi_2 \in D(p_j) \cap \HH_{\rm
fin}$ by taking linear combinations and then limits.
\end{proof}

\end{document}